\documentclass[letterpaper, 10 pt, conference]{ieeeconf}  

\IEEEoverridecommandlockouts                              
\usepackage{hyperref}
\usepackage{multicol} 

\usepackage[svgnames]{xcolor} 
\usepackage{algorithm}
\usepackage{algpseudocode}
\usepackage{times} 
\algnewcommand\algorithmicforeach{\textbf{for each}}
\algdef{S}[FOR]{ForEach}[1]{\algorithmicforeach\ #1\ \algorithmicdo}

\usepackage{graphicx} 
\graphicspath{{figures}} 
\usepackage{booktabs} 

\usepackage{amsfonts, amsmath, amsthm, amssymb} 
\usepackage{wrapfig} 
\usepackage{epstopdf}

\title{\LARGE \bf
Tuning of Drone PD Controller Parameters for Medical Supplies Delivery*\thanks{*This short paper is extracted from \cite{azin2021} for \href{https://sites.google.com/ualberta.ca/2021-icra-workshop/homeh.4sy4u1fliv7u}{ICRA 2021 Workshop on Impact of COVID-19 on Medical Robotics and Wearables Research
}.}}

\author{ \parbox{2 in}{\centering Azin Shamshirgaran*
        \thanks{* Corresponding author}\\
         Department of CSE \\ 
         University of California, Merced\\
         Merced, CA 95343\\
         {\tt\small ashamshirgaran@ucmerced.edu}}
         \hspace*{ 0.5 in}
         \parbox{2 in}{ \centering Hamed Javidi\\
         Department of EECS \\ 
         Cleveland State University\\
         Cleveland, OH 44115\\
         {\tt\small h.javidimostafapourboshroyeh\\@vikes.csuohio.edu}}
         \hspace*{ 0.5 in}
         \parbox{2 in}{ \centering Dan Simon\\
         Department of EECS \\ 
	Cleveland State University\\
	Cleveland, OH 44115\\
         {\tt\small d.j.simon@csuohio.edu}}
}

\begin{document}

\maketitle
\thispagestyle{empty}
\pagestyle{empty}

\begin{abstract}
During the COVID-19 pandemic and similar outbreaks in the future, drones can be set up to reduce human interaction for medical supplies delivery, which is crucial in times of pandemic. 
In this short paper, we introduce the use of two evolutionary algorithms for multi-objective optimization (MOO) and tuning the parameters of the PD controller of a drone to follow the 3D desired path. 
\end{abstract}
\section{INTRODUCTION}
Unmanned aerial vehicles (UAVs), such as drones and quadrotors, have gained significant attention during the last decade.  UAVs can be used in many fields, for instance, inspecting and exploring new environments, monitoring weather patterns to predict tsunamis and earthquakes, construction, monitoring gas and oil resources, and performing jobs in dangerous environments with the advantages of high robustness, reliability, stability, and low resource consumption~\cite{shamshirgaran2016dynamic}, \cite{azin2020}.

During the COVID-19 pandemic and similar outbreaks in the future, drones can be set up to improve the everyday lives of people. Drones are effective at reducing human interaction, which is crucial in times of pandemic. To reduce the risk of coronavirus infection, governments have asked and encouraged people to remain in their homes. But then, there should be a way to provide services and support for people in their homes.
Drones can be used for that purpose by facilitating contact-free interactions with healthcare professionals, such as transporting blood or urine samples, and delivering medical supplies like medicine or healthcare devices. During a pandemic, hospitals are potential vectors of contamination, so drones provide an efficient contact-free way to transport critical and necessary medical supplies.  
Although medical supply delivery has been achieved by the commercial company DJI~\cite{djitest}, there are still many challenges, some of which we focus on in this paper. 

 Considering system and environmental noise, many research studies have focused on developing a drone controller to maintain stability, to reach the defined objective, or to tune the controller's parameters. The most well-known controllers for drones are proportional-integral-derivative (PID) and proportional-derivative (PD). Since they are widely applied for drone control, there are a lot of studies on tuning the parameters of these controllers. PID control includes three adjustable gain parameters:
the proportional gain $K_p$, integral gain $K_i$ and derivative gain $K_d$. Algorithms proposed for this tuning problem mostly used aggregation-based multi-objective optimization, which uses a weighted sum of different cost functions to tune the controller parameters. These tuning algorithms require the arbitrary determination of weight coefficients and the main advantages of them are simplicity and computationally efficiency. 

Biogeography-based optimization~(BBO) and particle swarm optimization (PSO) are two popular optimization algorithms, so we focus on extending these two algorithms to MOO for drone control in this paper. PSO is based on candidate solutions sharing positions in solution space with each other. Each candidate solution, or particle, evolves its position in solution space based on the locations of other particles, until a desirable solution is found~\cite{Cai2007}. BBO is based on islands sharing (or migrating) suitable features, which represent independent variables in the problem solution \cite{simon2008biogeography}. Each island is considered as a possible solution for the problem. Islands gradually evolve by migrating other islands' features to become better habitats (i.e., better solutions) until a desirable solution is found. 

In this short paper, we use multi-objective optimization along with evolutionary algorithms to tune the parameters of the PD controller of a drone. The multi-objective function is based on the tracking error of the four states of the system. 
 \section{Dynamic Model of the Drone}
We used the Euler-Lagrange model to derive the equations of the drone \cite{azin2020}. The linear and angular position of the drone are defined in relation to the inertial reference frame $x$-$y$-$z$. The angular velocities $p$, $q$, $r$ are defined in relation to the body reference frame $x_B$-$y_B$-$z_B$. The pitch rotation of the drone around the $y$-axis is denoted by $\theta$, the roll rotation around the $x$-axis is denoted by $\phi$, and the yaw rotation around the $z$-axis is denoted by $\psi$. The center of mass of the drone is located at the origin of the body frame. Vector $\epsilon=[x, y, z]^T$ represents linear position, $\eta=[\phi, \theta, \psi]^T$ represents angular position, and $\nu=[p, q, r]^T$ represents angular velocity in the body frame. The drone contains four rotors which induce angular velocities $\omega_i$, torques $M_i$ and forces $f_i$. Thrust $T=f_1+f_2+f_3+f_4$ is created by the combined force in the $z$ axis, and torques $\tau=[\tau_\theta, \tau_\phi, \tau_\psi]^T$ are created in the body frame. Please refer to \cite{azin2020} and \cite{azin2021} for the equation used for the dynamic model.

\section{Aggregation Method}
\label{sec-Aggregation}
The main advantages of the aggregation method are simplicity and computational efficiency. 
The aggregated objective function based on the four most important criteria in the PD controller is defined as 
\begin{equation}
\label{eq:q9}
\begin{split}
  F(X)= \omega_1 F_1(X) + \omega_2 F_2(X) + \omega_3 F_3(X) + \omega_4 F_4(X)  \\
  \end{split}
\end{equation}
where $w_i$ is the weight for each individual objective function, which is set equal to 1. Each components of $F(X)$ is defined as
\begin{equation}
\label{eq:q8}
\begin{split}
    F_{i}(X) = \int_{t}|X_i-X_{i,d}| \, dt
  \end{split}
\end{equation}
where $i \in [1, 2, 3, 4]$ indexes the objective, the state $X_i \in \{\phi, \theta, \psi, z\}$, and the desired (reference) state $X_{i,d} \in \{\phi_d, \theta_d, \psi_d, z_d\}$ \cite{azin2021}.
 
\section{Simulation Results}
We evaluate the performance of multi objective BBO (MOBBO) and multi objective PSO (MOPSO) on the drone controller via computer simulation using MATLAB/Simulink.
We used the following parameters for MOBBO.\\
\begin{equation}
\label{eq:10}
\begin{split}
  p_s = 50,
  I_t = 30\\
  E_r = ( p_s  +1 - f_s) / ( p_s + 1); f_s\in[1,2,...,p_s]\\
  I_r= 1- E_r\\
  N_e = 2
  \end{split}
\end{equation}
where $p_s$, $I_t$, $E_r$, $I_r$ and $N_e$ are population size, iteration limit, emigration rate, immigration rate, and number of elites. $N_e$ is the number of the best solutions to keep from one generation to the next. For MOPSO,
\begin{equation}
\label{eq:11}
\begin{split}
  p_s = 50,
  I_t = 30\\
w=0.5,
w_d = 0.99 \\    
 c_1 = 2,
c_2 = 2
  \end{split}
\end{equation}
where $p_s$, $I_t$, $w$, $w_d$, $c_1$ and $c_2$ are population size, 
iteration limit, inertia weight, inertia weight damping ratio, personal learning coefficient and global learning coefficient.
The desired position $z_d$ of the drone is fixed at~$z_d = 0$, and the desired angular positions are fixed at
$\theta_d = \phi_d = \psi_d = 0$ with the initial  positions and Euler angles selected as $\begin{bmatrix}
x_0, & y_0, & z_0  \end{bmatrix}^T =\begin{bmatrix}
0, & 0, & -1  \end{bmatrix}^T$, and $\begin{bmatrix}
\theta_0, & \phi_0, & \psi_0  \end{bmatrix}^T = \begin{bmatrix}
-0.7, & -0.7, & -0.7  \end{bmatrix}^T$. The drone parameters in our simulations are shown in \cite{azin2020}.
 \subsection{Simulation Results of Aggregation Method}
\begin{figure}[b]
      \centering
      \includegraphics[scale=0.16]{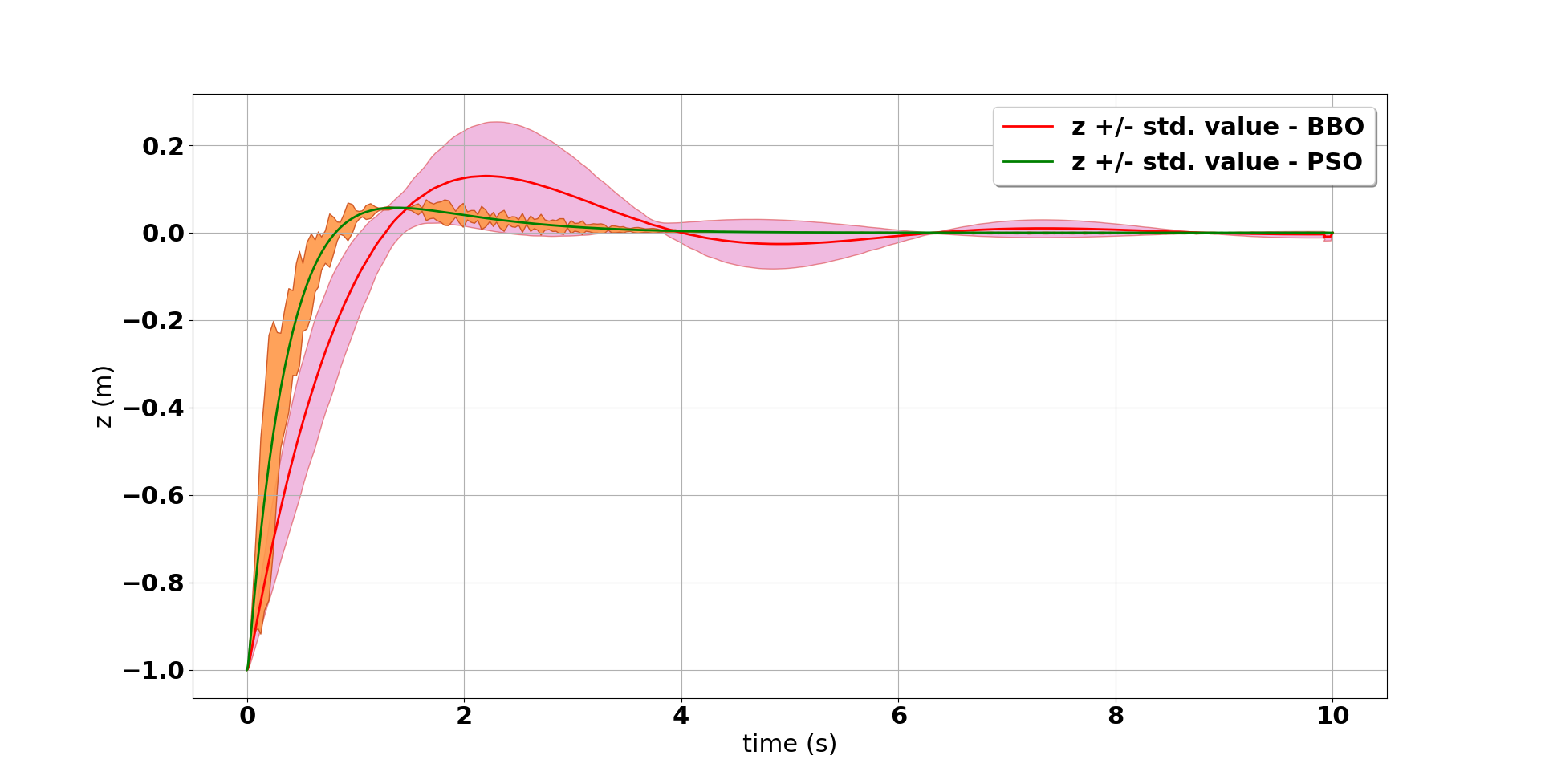}
      \caption{Mean and standard deviation (shaded) of $z$ of 5 trials of MOBBO and MOPSO}
      \label{p4}
\end{figure}

 In Table~\ref{tabel11}, the Min and Max columns show the search space bounds for the PD parameters, the PD column shows the values used for the conventional PD controller~\cite{luukkonen2011modelling}, and the PSO and BBO columns show the mean values of the PD parameter that PSO and BBO converged to after 30 iterations and 5 trials.
Figure~\ref{p4} shows the mean and standard deviation of state $z$ for both MOPSO and MOBBO with 0.05 and 0.1 m overshoot, respectively. MOPSO has a better rise time, about 1 sec, and a better settling time, about 3 sec. For more results and discussions refer to~\cite{azin2021}.

\begin{table}[t]
\caption{PD tuning comparison}
\label{tabel11}
\begin{center}
\begin{tabular}{ |c|c|c|c|c|c| } 
\hline
 & Min & Max & PD & PSO & BBO  \\
\hline
\hline
$K_{p_{\phi}}$ &0&20.0&6&14.015&19.7704\\
\hline
$K_{d_{\phi}}$ &0&10&1.75&10&9.6322\\
\hline
$K_{p_{\theta}}$ &0&10&6&2.7624&3.04\\
\hline
$K_{d_{\theta}}$ &0&10&1.75&10&9.6105\\
\hline
$K_{p_{\psi}}$ &0&10&6&6.4304&1.49\\
\hline
$K_{d_{\psi}}$&0&10&1.75&10&9.9945\\
\hline
$K_{p_{Z}}$&0&3& 1.5 & 3 & 2.8141 \\
\hline
$K_{d_{Z}}$& 0&3&2.5 & 2.7755 & 2.89\\
\hline
\end{tabular}
\end{center}
\end{table}

\section{CONCLUSIONS}
In this short paper, we used evolutionary algorithms to tune the parameters of a PD controller of a drone. The multi-objective function is defined based on the tracking errors of the four states of the system, which in turn help reduce settling time, overshoot, rise time and steady state error.   
 The results showed improved results compared to conventional PD control. 

\bibliographystyle{IEEEtran} 
\bibliography{azin} 
\end{document}